# Machine learning for precision psychiatry

Prof. Danilo Bzdok, M.D., Ph.D.[1,2,3] & Prof. Andreas Meyer-Lindenberg, M.D.[4,5]

1 Department of Psychiatry, Psychotherapy and Psychosomatics, RWTH Aachen University, 52072 Aachen, Germany
2 JARA-BRAIN, Jülich-Aachen Research Alliance, Germany
3 Parietal team, INRIA, Neurospin, bat 145, CEA Saclay, 91191 Gif-sur-Yvette, France
4 Department of Psychiatry and Psychotherapy, Central Institute of Mental Health, Medical Faculty Mannheim, University of Heidelberg, J5, 68159 Mannheim, Germany
5 Bernstein Center for Computational Neuroscience Heidelberg-Mannheim, Central Institute of Mental Health, Medical Faculty Mannheim, Heidelberg University, J5, 68159 Mannheim, Germany

**Abstract**

The nature of mental illness remains a conundrum. Traditional disease categories are increasingly suspected to mis-represent the causes underlying mental disturbance. Yet, psychiatrists and investigators now have an unprecedented opportunity to benefit from complex patterns in brain, behavior, and genes using methods from machine learning (e.g., support vector machines, modern neural-network algorithms, cross-validation procedures). Combining these analysis techniques with a wealth of data from consortia and repositories has the potential to advance a biologically grounded re-definition of major psychiatric disorders. Within the next 10-20 years, incoming patients could be stratified into distinct biological subgroups that cut across classical diagnostic boundaries. In a new era of evidence-based psychiatry tailored to single patients, objectively measurable endophenotypes could allow for individualized prediction of early diagnosis, treatment selection, and dosage adjustment to reduce the burden of disease. This primer aims to introduce clinicians and researchers to the opportunities and challenges in bringing machine intelligence into psychiatric practice.

**Introduction**

By relying on human-conceived diagnostic groups, psychiatric treatment is explored in scientific research, evaluated for effectiveness in clinical trials, and administered every day to suffering patients. These pervasively adopted diagnostic categories have been constructed from expert opinion and enshrined in the DSM-5 and ICD-10 manuals. Yet, it is becoming increasingly clear that the pathophysiology underlying such disease definitions is rather heterogeneous (1, 2). A clinically distinct mental disease is often not underpinned by an identical biology as far as we can detect by available neuroscientific instruments. This frustration can potentially be alleviated by identifying subgroups that exhibit predictable response to treatment. The aspiration to automatically segregate brain disorders into natural kinds will however necessitate new statistical and scientific approaches.

For decades, the dominant research paradigm to alleviate symptoms of psychiatric patients has followed an ideal chain of events: 1) initially neuroscience studies should identify new disease mechanisms (e.g., neurotransmitter pathways in animal models), 2) then novel treatments should be explored to target the discovered disease mechanisms (e.g., design and test candidate molecular compounds), and 3) finally the new treatment should be validated by clinical trials in large cohorts (e.g., randomized clinical drug trials). Despite numerous important advances, each of these three steps has encountered considerable difficulties.

First, the search for novel disease mechanisms has yielded sobering results in many mental diseases (3, 4). A contributing reason may be that animal models of psychiatric disorders typically immitate some properties of a given brain disorder without emulating the disorder itself (5). Second, over recent years, major players in the pharmaceutical industry have been abandoning their investment programs for psychotropic drug discovery (6). This trend can be observed in the USA (Pfizer and Merck), Switzerland (Novartis), France (Sanofi), and UK (GlaxoSmithKline and AstraZeneca). As a consequence, innovative drug treatments become even less likely to enter the market in the next 10 to 20 years. Third, it is often only in later stages of the numbingly expensive clinical trials that many psychotropic drug candidates turn out to be ineffective or not usable in humans (7, 8). In wanting to supplement this traditional paradigm in psychiatric research, modern *machine learning* approaches probably lend themselves particularly well to ultimately improve the well-being of psychiatric patients.

Machine learning uses quantitative models to induce general principles underlying a series of observations without explicit instructions (9, 10). Such algorithmic methods are characterized by 1) making few a-priori assumptions, 2) allowing the data to "speak for themselves", and 3) the ability to mine structured knowledge from extensive data. Its members include *supervised methods*, such as support vector machines and neural-network algorithms, specialized for best-possible outcome prediction as well as *unsupervised methods*, such as algorithms for data clustering and dimensionality reduction, effective at discovering novel statistical configurations in data (see Table 1 for technical terms). The recent coincidence of increasing data availability, improving computing power, and cheaper data storage has encouraged an ongoing surge in research and applications of machine learning technologies roughly since the turn of the century, although a majority of these tools had already emerged in the second half of the 20th century (11). Here, input versus output variables for quantitative modeling are called "features" versus "target variables" (12), familiar to many as independent versus dependent variables. As another special property, inferring new domain knowledge routinely takes the empirical form of extrapolating patterns by successfully predicting information about previously unseen observations in *cross-validation procedures* (13). In the following we will argue that machine learning is predisposed to address many upcoming challenges in the era of precision medicine in psychiatry.

**Opportunities**

Current drug treatment choices are only successful in roughly every second patient (14), and similar considerations apply to psychotherapy. An alternative research strategy that does not depend on full understanding of complex disease mechanisms may therefore be cheaper and incur shorter delays between bench and beside (15). Indeed, the psychiatrist's choice of the best-possible treatment option often does not depend on knowledge of what has caused or maintains the mental disease of a given patient. Systematically benchmarking the predictability of clinical quantities in *single patients* could faster improve clinical symptoms and reduce subjective suffering in many mental diseases. Even moderately successful predictive models can be highly useful in clinical practice. This is because of the unfortunate normal case of trial-and-error treatment with psychotropic drugs and other types of treatment for many mental diseases (16). While the traditional research goal was to introduce *novel* treatment options that benefits some *majority* of the respective clinical group, an attractive alternative research goal is to improve the choice of *existing* treatment options by predicting their effectiveness in *single* patients.

Psychiatric research has long been at odds with conventions in clinical practice. A disconnect is apparent between medical training ingrained in diagnostic categories and qualitative medical care idiosyncratic to particular individuals. Even if clinical guidelines are typically backed up by scientific results on group effects, psychiatrists frequently act on a patient-by-patient basis on the ward. In contrast to other medical specialties, treatment choice in psychiatry is to a larger extent guided by symptoms and phenomenology of a particular patient, rather than exclusively dictated by the patient's diagnosis. However, scarce evidence exists to reduce the uncertainty of which treatment will benefit which patient. No objective biomarkers, be they derived from blood, brain, or genes, are currently available for treatment outcome prediction in psychiatric patients, incurring social crisis, life-quality costs, and socioeconomic burden. More and more studies now indicate that a specific drug or psychotherapy treatment can be successful in a certain patient subgroup and unsuccessful in another patient subgroup labeled with an identical diagnosis (see here for an overview: 17). In a seminal machine learning study, discovered patient subgroup could indeed predict which patient would profit from brain-stimulation treatment (18). This questions the primacy of drawing conclusions on the group-level and opens the possibility of building objective algorithmic frameworks with individual treatment-response prediction across a diversity of psychiatric conditions.

Machine learning offers a set of tools that are ideally suited to achieve individual-level clinical predictions. Predictive models are conceptually positioned themselves between clinical symptoms and genetic risk variants, which has the translational potential to refine clinical management by early diagnosis and disease stratification, selection between drug treatments, treatment adjustment, and prognosis for psychiatric care tailored to each individual (18). Learning algorithms can thus be readily applied in single patients to predict inherently valid and immediately useful clinical objects (19), such as choosing drug dosage. There is a number of reasons why many machine learning methods are naturally applicable for prospective clinical predictions on the single-subject level, whereas the currently most widespred statistical methods may be more tuned to group-level analysis:

<u>i) Focus on prediction:</u> Machine learning methods have always had a strong focus on *prediction* as a metric of statistical quality (12). Support vector machines, neural-network algorithms, and many other predictive models are hence readily able to estimate an outcome from only one data point, such as when querying answers from behavioral, neural, or genetic measurements of a single patient (20, 21). In contrast, classical statistical methods are often used in medical research to *explain variance* of and *formally test* for group effects. ANOVA, t-test, and many other commonly used tools underlying the notion of statistical significance have a less obvious ability for judgements on one specific individual in a group. Thus, routines of machine learning and classical statistics serve rather distinct statistical purposes. The two statistical cultures perform different types of formal assessment for successful extrapolation of an effect beyond the data at hand that are rooted in different mathematical contexts. As an important practical consequence, machine learning and classical statistics do not judge data on the same aspects of evidence: An observed effect assessed to be statistically significant by a p-value does not in all cases yield high prediction accuracy in new, independent data, and vice versa (10, 21-23).

ii) Empirical model evaluation: By quantifying the prediction success in new individuals (so-called *out-of-sample* estimates) many machine learning approaches naturally adopt a prospective viewpoint and can directly yield a notion of clinical relevance. Instead, classical approaches based on null-hypothesis testing often take a retrospective flavor as they usually revolve around finding statistical effects in the dataset at hand (so-called *in-sample* estimates) based on a pre-specified modeling assumptions, typically without explicitly evaluating some fitted models on unseen or future data points (22). Hence, techniques for out-of-sample generalization common in machine learning are likely candidates for enabling a future of personalized psychiatry. This is because predictive models can be applied to and obtain answers from a single patient.

iii) Two-step procedures: Similarly, traditional null-hypothesis testing takes the form of a *one-step procedure*. That is, the whole dataset is routinely used to produce a p-value or effect size measure in a single process. An obtained p-value or effect size can itself not be used to judge other data in some later step. In contrast, machine learning models are typically evaluated by cross-validation procedures as a gold standard to quantify the capacity of a learning algorithm to extrapolate beyond the dataset at hand (12). In a *two-step procedure,* a learning algorithm is fitted on a bigger amount of available data (so-called "training data") and the ensuing "trained" learning model is empirically evaluated by application to a smaller amount of new data (so-called "test data"). This two-step nature of machine learning workflows lends itself particularly well to, first, extract structured knowledge in large openly available or hospital-provided datasets. Second, the ensuing trained predictive models can be shared collaboratively as a research product (24) and be applied with little effort in a possibly large number of individual patients in various mental health contexts.

iv) Suited to observational data: Many methods from classical statistics have probably been devised for *experimental data* that are acquired in a context where a set of target variables has been systematically manipulated by the investigator (e.g., randomized clinical trials with placebo group and active treatment group). Precision medicine in psychiatry is however likely to exploit especially *observational data* (e.g., blood and metabolic samples, movement and sleeping patterns, EEG, brain scans, and genetic variants) that were acquired without a carefully controlled influence in an experimental setup and to which machine learning tools may be more closely tuned (e.g., 18, 25, 26).

v) Handle many outcomes: Machine learning is also a pertinent choice for comparisons between possible diagnoses and other multi-outcome settings. Classical significance testing is traditionally used to decide between two possible outcomes, expressed in the *null* and *alternative hypothesis*, by considering the probability of obtaining an equal or more extreme effect in the data under the null hypothesis (27). This is often used in group analysis to formally determine a scientifically relevant difference between healthy subjects (i.e., typically corresponding to the null hypothesis) and psychiatric patients as defined by a DSM or ICD category (i.e., typically the alternative hypothesis), or when comparing a placebo treatment (i.e., null hypothesis) against a new treatment (i.e., alternative hypothesis). In everyday practice on the psychiatric wards, the more challenging question is probably not whether a patient has a mental disease or not but the differential diagnosis between a number of likely disease categories --- the *trans-diagnostic* setting. Analogously, whether a patient needs treatment or not is routinely an easier clinical decision than choosing between a number of competing treatment options. Treatment response prediction is rarely a binary yes-or-no decision and requires consideration of several treatment options in the same statistical analysis.

Machine learning is well suited to this goal in the form of *multi-class prediction* and *multi-task learning (26, 28, 29),* unlike many approaches for statistical significance and tests of group differences. Most machine learning approaches that are applicable when aiming to distinguish two groups or two treatment options can be readily extended to considering a range of possible outcomes. For instance, quantitative brain measurements from one patient can be fed into prediction models to infer a probabilistic stratification not only over several differential diagnoses and over many candidate treatment options, but also risk outcomes, or possible long-term clinical prognoses (e.g., full recovery versus partial residuals versus severe chronic illness). Additionally, applying learning algorithms to compare patients versus controls does not allow evaluating how specific an achieved prediction is for the given psychiatric group (24). Besides the advantage of replacing artificial, mutually exclusive dichotomies by predicting several outputs in concert, the prediction accuracy does also often improve when statistical strength can be shared between the variation in the data associated with the different outcomes (30). In sum, there are clear incentives and readily applicable statistical tools to go beyond group-level comparisons à la normal versus diseased (31, 32). Importantly, machine learning is naturally suited for choosing between a potentially large number of possibilities in a single patient and ranking the possibilities according to pertinence. This is the case in the trans-diagnostic setting where the pertinence of several psychiatric diagnoses needs to be predicted simultaneously by one statistical model.

vi) Explore manifolds in complex data: Besides the intricacies of considering several diagnostic categories at once, the diagnostic categories themselves have repeatedly been called into question due to their lack of neurobiological validity and clinical predictability (1, 2). The disease definitions cataloged in the DSM and ICD manuals do not always align well with new behavioral, neuroscientific, and genetic evidence (Fig. 1). Psychiatric disorders have been defined in the DSM and ICD with a focus on ensuring effective communication of diagnoses (i.e., inter-rater reliability) between clinicians rather than the goal to capture natural kinds in biological reality (1). Autism, schizophrenia, and an increasing number of other psychiatric diseases are suspected to be *spectrum disorders* --- heterogeneous etiological and pathophyisiological factors being summarized under the same umbrella term (33, 34). This conceptualization is also more compatible with a smooth transition between healthy and psychiatrically diagnosed individuals. Machine learning offers a rich variety of tools that readily lend themselves to endophenotype modeling.

Among many clinicians and researchers, there is hence a growing wish to supplement discrete disease definitions in form of categories with a continuous, dimensional symptom system. To satisfy the need to cut across diagnostic boundaries, the Research Domain Criteria (RDoC) initiative (35) has been launched as a translational program to elucidate the hidden structure underlying psychopathology. By synergistic integration of self-reports, neuropsychological tests, brain measurements, and genetic profiles, RDoC wants to "better understand basic dimensions of functioning [...] from normal to abnormal" (National Institute of Mental Health) without relying on presupposed disease definitions. The discovered fundamental dimensions of behavior and its disturbances are expected to motivate new research approaches aimed at re-formating psychiatric nosology. RDoC thus recommends going from scientific evidence to organically deriving new disease factors. This framework thus contrasts the dominant agenda in psychiatric research that goes from disease categories defined based on the DSM and ICD to generating scientific evidence. The RDoC approach is conceptually compatible with the fact that psychiatric patients exhibit clusters of psychopathological symptoms and that many symptoms are shared among, rather than unique to different psychiatric disorders. RDoC is also naturally compatible with the accumulating evidence that risk alleles are partly shared between psychiatric disorders (36), while different sets of risk alleles can lead to an identical psychiatric phenotype (37).

Of note, this renewed focus on fundamental building blocks of mental disturbance finds a direct correspondence in the long-standing focus of the machine learning community on *representation* *learning* for discovering hidden structure in complex data (38). In particular, the multi-dimensional conception underlying the RDoC initiative is reminiscent of the notion of *manifolds* that is common in the machine learning field (9). In a setting with possibly many high-resolution measurements (brain scans, sequenced genome, etc.), a manifold describes coherent low-dimensional directions of relevant variation in the data. Here, members of a coherent class would be expressed as "a set of points associated with a neighborhood around each point" (9). In psychiatry, the manifold notion corresponds to the hope that the nature of psychiatric disorders and their complex relationships could be described effectively in a small number of hidden dimensions: each a distinct direction of variation in heterogeneous data sources. Variation captured across behavioral, experiential, neural, and genetic measurements with tens of thousands of input variables can hopefully be effectively expressed along a manifold that concerns only a much smaller number of yet-to-be-discovered disease dimensions.

Indeed, machine learning has a rich legacy of algorithm developments that can now be repurposed to automatically extract manifolds from data describing behavior, life experience, brain, or genetics. Representation learning algorithms operate on the assumption that the measured data have been generated by a set of underlying constituent factors. Unfortunately, however, many currently used traditional clustering algorithms, such as hierarchical and k-means clustering, share the inconvenience to assign each individual exclusively to one cluster (so-called "winner-takes-all" assumption). This biologically and clinically implausible statistical assumptions can be relaxed by recourse to *latent factor models* (9, ch. 13), including latent Dirichlet allocation (25), autoencoders (39), and many other dictionary learning procedures. Latent factor models can readily uncover an underlying manifold of hidden directions of variation by assigning each individual to *each of the clusters to different degrees*. Technically, the same manifold dimension, reflecting a distinct disease process, is thus allowed to contribute in nuanced ways to several psychiatric disorders with clinical pictures as diverse as schizophrenia, autism, and bipolar disorder. Thus, given the prevailing lack of objective markers in psychiatry, there is merit in revealing, formalizing, and clinically exploiting currently unknown inter-individual variation.

For a long time, knowledge generation in basic neuroscience and clinical decision-making in psychiatry has been dominated by classical statistics with formal tests for group differences in frequently small samples. Machine learning methods however better appeal to the ambitions of precision psychiatry because they can directly translate complex pattern discovery in "big data" into clinical relevance. For most learning algorithms, it is standard practice to estimate the generalization performance to other samples by empirically cross-validating the trained algorithms on fresh data; in this case, *individual subjects*. This stands in stark contrast to classical statistical inference that seeks to reject the null hypothesis by considering the entirety of a data sample (27); in this case, *all available subjects*. In the inferential framework, the desired relevance of a statistical relationship in the general population is ensured by formal mathematical proofs and is not commonly ascertained by explicit model evaluation on independent data (10, 27). It will be of growing importance that many traditionally used tools testing for statistical significance are categorically incapable of affording individual-level model predictions --- a cornerstone for personalized medicine.

**Challenges**

Many reasons speak in favor of extending machine learning techniques to psychiatric research and practice. The flipside of the coin is that there are still a number of obstacles to overcome:

i) Reproducibility: Prototyping, iteratively improving, and benchmarking machine learning pipelines involves many complicated and inter-dependent choices. Such multistep workflows are becoming challenging to fine-tune manually; hence requiring computer programming skills that will probably be in increasing demand. The flexibility of machine learning pipelines is raising the concern that obtained findings might less reliably replicate in later studies (40, p. 185). Successful deployment of predictive models on the clinical ward may require seamless exchange of predictive models in the research community. Once acceptance for model sharing is gained and effective platforms are set up, the prediction performance of a model can undergo further validation at different levels (24): The final predictive models should pass the prospective test of new subjects in other research laboratories, persevere across data acquisition means (e.g., brain scanners from Siemens, Philips, and GE), across geographic locations (e.g., USA, Europe, and Asia) and across populations (e.g., same mental disorder with different comorbidity profiles), as well as for different success metrics (e.g., sensitivity and specificity) and different clinical settings (e.g., rural practitioner versus university hospital). In the future, a predictive model could improve step by step through annotation and modification in various laboratories in a cost-effective cumulative collaboration (24). Moreover, clinical biomarkers derived from genetics or brain-imaging will probably be accredited through randomized clinical trials, like other treatment candidates for psychiatry.

ii) Data availability: The primary limitation for deploying state-of-the-art machine learning to personalize psychiatric care is probably the size of today's datasets (i.e., number of subjects) and their insufficient phenotypic descriptions (e.g., medical history, comorbidities, progression in symptoms, treatment and response). A recent statement claimed that the "field of mental health captures arguably the largest amount of data of any medical specialty" (41). However, compared to some non-medical research domains, psychiatric research is still far from the sample sizes of n > 1,000,000 where the predictive power of the currently most successful, yet most data-demanding models has been showcased (9). Thousands of observations per category (e.g., brain scans or genetic profiles of responders to one specific drug treatment) are typically required to reach reasonable prediction accuracy with so-called "deep neural-networks" (9), while these models repeatedly achieved supra-human performance with millions of observations. Even ongoing prospective population studies, such as the burgeoning UK Biobank, aim at brain scans from "only" n = 100,000 participants to be completed in 2022. Similarly, genome-wide analyses on a specific phenotype have rarely reached sample sizes of more than n = 100,000 participants. Besides limited data quantity, exploiting emerging machine learning technologies is hindered by insufficient specificity and granularity of the participants' behavioral information. On the one hand, many phenotypes of interest do not vary enough across subjects in general-purpose datasets. On the other hand, deploying current machine learning for successful subject-level prediction of practically useful clinical endpoints will probably depend on datasets with rich and meticulously acquired patient documentation from various time points. Soon, high-throughput predictive models may be satisfied by the increasing digitalization of everyday life by ubiquitous electronic sensors (42).

iii) Data management: Over the last 10 years, growing sample sizes were enabled by national and international consortia to accumulate, curate, and distribute data across research groups, exemplified by the Autism Brain Imaging Data Exchange (ABIDE) and Alzheimer's Disease Neuroimaging Initiative (ADNI). An important prerequisite resides in the willingness to embrace the values and habits of the open-science movement (43), potentially prompting re-calibration between society's good and citizens' rights. While some investigators deem restricted access to research data unethical, data sharing also invokes privacy concerns (44). For instance, a participant's face could be recognized from anatomical brain scans, but this sensible information can be automatically removed by de-facing techniques. Similarly, the Human Connectome Project (HCP) has adopted a multi-tier access policy: Most data are openly available from all subjects, whereas access to their sensitive information is controlled. As another trend, the same dataset can be disseminated in different variants: from sharing only 3D coordinates of neural activity changes, over precomputed aggregate data as statistical summaries to preprocessed or full raw data (21, 44). Further, retrieving an extensive dataset requires considerable transmittion bandwidth and may be infeasible or unallowed. The ENIGMA consortium has demonstrated large-scale analysis of data from >1000 subjects stored in distant research centers using distributed model fitting (45). Emerging means for decentralized model building can enable large federated statistical analysis precluding the need for data pooling in a single location. Consequently, agreement on machine-readable data structures will grow in importance (46).

iv) <u>Heterogeneous and incomplete data:</u> A first generation of data initiatives (e.g., ABIDE, ADNI, ENIGMA) were retrospective collections of independently acquired data from different clinical centers. Such data repositories typically vary in data quality, acquisition parameters, hardware and software versions, preprocessing procedures, artifact occurrence and quality control, used psychological assessments and clinical questionnaires, missing data, and population aspects. Across-site heterogeneity may explain why, counter-intuitively, predictive model performance have been repeatedly reported to decrease as the available neuroscience data increases (24). A second generation of data initiatives (e.g., HCP and UK Biobank) realized prospective collections that early convened on how to standardize data acquisition. Ensuing repositories offer higher data comparability due to efforts including calibrated acquisition conditions, staff training, or traveling experts. Of note, homogenizing data acquisition and analysis can maximize group differences and alleviate confound problems, whereas homogenizing population samples may not be optimal in all cases. A majority of clinical studies recruits patient samples based on tight inclusion and exclusion criteria. Such handpicked subject groups (e.g., excluding frequent comorbidities of a disease) are often less representative of the broader population. Derived predictive models may not perform well on the genuinely heterogeneous patients on the psychiatric ward. Hence, a balance must be found between conservative manual selection of samples with convincing model performance and liberal samples more representative of clinical reality.

v) <u>Longitudinal data:</u> Retrospective data collections typically lend themselves more to cross-sectional analysis, while prospectively collected data are often suitable for longitudinal analysis. Most machine learning approaches and their clinical applications currently focus on cross-sectional findings. Computational psychiatry research may bear a blind spot regarding disease trajectories and longer-term health outcomes (17), despite the time-varying nature of many mental disorders. A promising avenue to accumulate massive longitudinal data may be offered by technical devices carried by subjects (42), a rationale that has already energized many commercial health companies. For instance, voice data from smartphones and connected technologies could enable early detection of healthcare events, such as thought disorders, depressive episodes, or suicide attempts. More generally, digital sensors are becoming ubiquitous and are well positioned to continuously monitor diverse behaviors, including sleep patterns, communication habits, as well as gait and geographical movement. This may enable evolving machine learning models that capture which factors contribute to a patient's symptom dynamics across time. In addition to privacy concerns, ditigital sensors are subject to constant change in hardware and software, which increases heterogeneity in the acquired longitudinal data.

vi) <u>Confounding:</u> Accumulating observational human data is often cheaper and easier, while lack of experimental protocols exacerbates control of confounding influences (47). Essentially, the prediction performance becomes inflated if the training data used for model fitting and the testing data are somehow mutually dependent, even if contaminated in subtle ways. Researchers are challenged to identify and account for influences unintentionally contributing to high prediction accuracies, including age, gender, culture, ethnicity, smoking, caffeine, movement effects (e.g., eye blinks), physiological noise (e.g., respiration and heart beat), and medication use. Confounding gets more complicated in continuously updated models: A biased predictive model could lead to clinical action that further increases the existing bias --- driving a feedback loop with self-fulfilling prophecies (48). Sociologically, investigators are also often restricted to hospitalized population samples. Bias may inadvertently arise because clinical research typically recruits subjects with exposure to psychiatric institutions, rather than never-diagnosed individuals with mental problems. For instance, high-functioning, subclinical schizophrenics, never in contact with a psychiatrist, may systematically evade research efforts. Finally, some confounding issues might be revealed and alleviated by shifting towards more nuanced prediction in a multi-outcome settings, such as simultaneously distinguishing schizophrenia, schizoaffective disorder, autism, bipolar and personality disorders.

**Conclusion**

The devastating collateral damage and soaring costs of psychiatric disease prompt a global challenge for our societies (49). Whether or not personalized medicine can be realized to enhance psychiatric care is largely a statistical question at its heart. For many decades, *the group* has served mental health investigators as the primary working unit. Facilitated acquisition of always more detailed and diverse information on psychiatric patients is now bringing another working unit within reach --- *the single patient*. Rather than preassuming existence of disease categories and formally verifying prespecified neurobiological hypotheses, an increasingly attractive alternative goal is to let the data guide the investigation. Following the new data richness and changing research questions, some long-trusted statistical methods may be superseded as the best tool in the box.

Machine learning offers a statistical culture that can readily appreciate the smooth transition between well-being and illness as well as the foggy boundaries between disease categories. Learning algorithms hold promise for the biologically grounded reconstruction of psychiatric disease descriptions by uncovering and leveraging inter-individual variation across behavior, experience, brain, and genetics. Data-derived disease manifolds can transcend the traditional disease categories portrayed in the DSM and ICD that today govern treatment choice and prognosis (24). The statistical properties of learning algorithms could thus enable clinical translation of empirically justified single-patient prediction in a fast, cost-effective, and pragmatic manner. Patient-level predictive analytics might help psychiatry to move from strong reliance on symptom phenomenology to catch up with biology-centered decision-making in other medical specialities (50).

From a larger perspective, mental health researchers struggle to verbalize mechanistic hypotheses for psychiatric disorders at the most pertinent abstraction level, ranging from molecular histone-tail methylation in the cell nucleus to urbanization trends in society as a whole. This epistemological challenge highlights more human-independent pattern learning algorithms as an under-exploited research avenue. Learning algorithms can automatically identify disease-specific biological aspects that achieve intrinsically valid and immediately useful clinical predictions. Ultimately, by allying with recent statistical technologies we may more likely impact mental disease that arises at the interplay between genetic endowment and life experience --- both of which are unique to each individual.

**Figure 1: Current challenges for precision medicine in psychiatry and possible solutions from machine learning**

*A)* Basic and clinical psychiatric research frequently investigates a given patient population by group comparison against the healthy population, possibly creating artificial dichotomies. Many machine learning approaches can naturally compare observations from a number of groups in the same statistical estimation (i.e., multi-class prediction; see also Table 1). *B)* The diagnostic categories in the ICD and DSM manuals were designed to reliably describe symptom phenomenology and are frequently incongruent with new behavioral, neural, and genetic research evidence. Machine learning methods can automatically extract currently unknown patterns of variation in individuals (i.e., manifold) simultaneously from heterogeneous data that cut across traditional diagnoses. *C)* Assigning diagnostic categories to patients ignores that different pathophysiological mechanisms (i.e., endophenotypes) can contribute to the same clinical picture. Instead of relying on categorical assignments, biologically defined subgroups can be described by continuous contributions of several disease processes in graded degrees (i.e., latent factor models). *D)* Psychiatric care often resorts to trial and error. Predictive models can improve patient care by earlier detection, treatment selection and adjustment, and inference of disease trajectory. After a machine learning algorithm has been trained in extensive data (i.e., in-sample), the trained predictive model can be used for personalized prediction without database access (i.e., out-of-sample). Reprinted with permission and modified from (2).

**Table 1: Overview of mentioned machine learning techniques** (in order of appearance)

| Notion | Purpose |
|---|---|
| Supervised learning | Models that predict a discrete outcome (e.g., healthy group versus control group) or continuous outcome (e.g., disease severity degrees) from measures of behavior (e.g., questionnaire), brain (e.g., neural activity), or genetics (e.g., single nucleotide polymorphisms)<br>Data have the form: features $X$ ($n$ subjects x $p$ variables) and target variable $y$ (one entry for each subject)<br>Example: Estimate patient prognosis based on genetic profile. |
| Unsupervised learning | Models that discover structure that is coherently present in the $p$ variables across subjects.<br>Data have the form: features $X$ ($n$ subjects x $p$ variables), but no target variable $y$.<br>Example: Reveal biological disease subgroups in patients based on genetic profile. |
| Clustering | A class of unsupervised methods that uses a certain criterion to segregate a set of elements into a number of groups according to their measured similarity. Many clustering models perform hard assignments: the groups are non-overlapping, with each element associated with only one group (i.e., 'winner-takes-all' assumption).<br>Example: k-means clustering, hierarchical clustering, spectral clustering, density-based spatial clustering (DBSCAN) |
| Dimensionality reduction | Re-expressing a set of observations in terms of their underlying essential patterns. |
| Support vector machines | A supervised model that performs prediction based on identifying observations in the data that are typical for the categories to be distinguished. |
| Neural-network algorithms | A supervised model that performs prediction based on a non-linear, multi-layer variant of linear regression. "Deep" neural-network are a modern version with a higher number of non-linear processing layers. |
| Cross validation | A two-step procedure used as the de-facto standard to estimate the capacity of a pattern learning model to extrapolate to future data samples. First, the predictive model is fitted on training data and, second, its generalization performance is evaluated on test data (out-of-sample). The process is repeated for different splits of the data (often 5 or 10 times). |
| In-sample estimate | Prediction performance measured in the same data that was also used to fit the model.<br>Example: Most applications of linear regression in biomedical research exclusively consider in-sample estimates, without considering out-of-sample estimates. |
| Out-of-sample estimate | Prediction performance measured in new data that was not used to fit a model.<br>Example: In machine learning, it is the core metric of how successful extrapolation of a derived pattern to new, independent data is quantified. |
| Training data | A model is fitted to identify a certain pattern from a larger part of the available data. |
| Test data | An already fitted model is used for prediction in a smaller part of the available data. |
| Multi-class learning | Applying a supervised model to predict an outcome $y$ that denotes more than two (possibly hundreds of) categories.<br>Example: Model predicts best among (many) more than two drug treatment options. |
| Multi-task learning | Applying a supervised model to simultaneously predict several outcomes $y_1, y_2, ..., y_m$.<br>Example: Model uses the same brain scans to conjointly predict drug treatment options, candidate diagnoses, and disease trajectories. |
| Manifolds | The statistical goal to reveal and express distinct factors that collectively underlying a set of observations.<br>Example: Everyday objects are manifolds in a 3D space (e.g., a flower), although there is a variety of perspectives from which humans can gather and contemplate information about an object, including vision, audition, touch, smell, and many others. |
| Representation learning | Applying models that can automatically extract hidden manifolds from data. |
| Latent factor modelling | A class of unsupervised methods that use a certain criterion to stratify a set of elements with their respective relationship to a number of hidden components of variation so as to maximize between-component dissimilarity. Many latent factor models perform soft assignments: component of variations are overlapping, with each element associated to each component to a certain extent (i.e., no 'winner-takes-all' assumption).<br>Example: latent Dirichlet allocation, autoencoders, nonnegative matrix factorization (NMF), isomap, t-distributed stochastic neighbor embedding (t-SNE) |
| K-means clustering | A popular clustering model that partitions the $p$ variables of $X$ into $k$ non-overlapping groups.<br>Example: Use genetic information to group mammels into human and non-human primates ($k = 2$). |
| Hierarchical clustering | A popular clustering model that successively builds a nested tree where the $p$ variables of $X$ are portioned into $k$ always more fine-grained non-overlapping groups. All clustering solutions from $k = 1$ group to $k = n$ groups are often computed.<br>Example: Use genetic information to group mammels ($k = 1$) into human and non-human primates ($k = 2$), which are then grouped into humans, apes, and monkeys ($k = 3$), and so forth. |
| Latent Dirichlet allocation | A latent factor model that stratifies count-like data into overlapping components of variation.<br>Example: Extract coherent combinations of number of times (positive discrete numbers) words occurred during an unstructured clinical interview. |
| Autoencoders | A latent factor model that stratifies continuous data into overlapping components of variation. |

|  | Example: Extract coherent combinations of item scores from structure clinical questionnaire (positive/negative non-discrete numbers). |
|---|---|
| Dictionary learning | Super-class of many clustering models and latent factor models. |

**Acknowledgments**

We thank Bertrand Thirion (INRIA), Teresa Karrer (RWTH), Julius Kernbach (RWTH), Cedric Xia (UPENN) and Efstathios Gennatas (UPENN) for valuable comments on a previous version of the manuscript.

The authors declare no conflict of interest.